\documentclass[12pt, a4paper]{article}
\usepackage{graphicx}
\usepackage[font=footnotesize]{caption}
\usepackage[hscale=0.8,vscale=0.83]{geometry}
\usepackage{amsfonts}

\begin{document}

\title{The black hole information loss puzzle, matter-gravity entanglement entropy and the second law}

\author{Bernard S. Kay$^*$  \medskip \\  {\small \emph{Department of Mathematics, University of York, York YO10 5DD, UK}} \smallskip \\ 
\small{$^*${\tt bernard.kay@york.ac.uk}}}

\date{}

\maketitle

\begin{abstract} 
\noindent
Since Hawking's 1974 discovery, we expect that a black hole formed by collapse will emit radiation and eventually disappear.  Closely related to the information loss puzzle is the challenge to define an objective notion of physical entropy which increases throughout this process in a way consistent with unitarity.  In recent years, this has been addressed with certain notions of coarse grained entropy.  We have suggested instead that physical entropy should be identified with matter-gravity entanglement entropy and that this may offer an explanation of entropy increase both for the black hole collapse and evaporation system and also for other closed unitarily evolving systems, notably the universe as a whole.   For this to work, it would have to be that the matter-gravity entanglement entropy of the late-time state of black hole evaporation is larger than the entropy of the freshly formed black hole.   We argue here that this is possibly the case due to (usually neglected) photon-graviton interactions.   If black hole evaporation is slowed down by putting the black hole in a slightly permeable box, we give plausibility arguments that the radiation remaining after a large black hole has evaporated will (be pure and) mainly consist of roughly equal numbers of photons and gravitons entangled with one another -- with a photon-graviton entanglement entropy possibly greater than the entropy of the freshly formed black hole.   It also seems possible that, even in the absence of such a box, the matter-gravity entanglement entropy might still increase and the late-time state again be a pure state of (predominantly) photons highly entangled with soft gravitons that the Hawking-emitted photons themselves had radiated.    More work is needed to find out if it is indeed so.
\end{abstract}

\bigskip

\bigskip

\noindent
After the (theoretical) discovery of black holes \cite{LesHouches} and prior to Hawking's (theoretical) discovery \cite{Hawking75} that black holes emit thermal radiation, it had seemed that it would be possible to violate the second law of thermodynamics by simply depositing a package of entropic matter into a black hole.  Bekenstein had pointed out \cite{Bekenstein} that the second law could be saved if a black hole itself had an entropy and argued that this entropy should be proportional to its area.   Hawking's discovery showed that this was indeed the case and (in units with $G=\hbar=c=k=1$) fixed the constant of proportionality to be $1/4$.   Thus an important puzzle was solved.   However Hawking's discovery also led us to question whether we had ever actually properly understood the second law.   One version of the second law is the statement that the entropy of a closed system increases monotonically with time.   And, in a quantum theoretic description based on Hilbert spaces and time-evolving density operators, there is a natural quantity \cite{vonN} with the properties one expects of an entropy, namely the \emph{von Neumann entropy}, $S^{vN}$ (sometimes called the `fine-grained' entropy) defined, for any density operator, $\rho$, to be $-{\rm tr}\rho\log\rho$.   And yet, one must face the fact that the following three assumptions

\medskip

\begin{itemize}

\item[(1)] Time evolution is unitary

\item[(2)] Physical entropy increases 

\item[(3)] Physical entropy is the von Neumann entropy of the total state

\end{itemize}

\medskip

\noindent
are in contradiction with one another because von Neumann entropy is a unitary invariant.  Let us call this the \emph{second law puzzle}.

Prior to Hawking's discovery, this had seemed not necessarily to be a problem because one could take the view that the physical entropy of a closed system is not its von Neumann entropy but rather some sort of coarse-grained entropy \cite{Reich,Davies}.   But, notoriously, coarse-graining is a subjective procedure \cite{Reich,Davies, Penrose79} (it depends on judgments about what states we, as physicists, can distinguish\footnote{\label{ftntCount} A particular form of quantum coarse-graining -- which is often applied to states which are energy eigenstates or close to energy eigenstates -- is to define the entropy of a (pure) state with (approximate) energy $E$ to be the logarithm of the number of energy eigenstates with energy in a small range about $E$ (or of the number with energy less than $E$) or, in the case of a highly degenerate energy level, just the number of energy eigenstates in that level.  This is often called the method of computing entropy by `counting states'.  This is subjective in that it relies on the judgment that we can't distinguish the given state from other states in that interval, or other states of the same energy. (In the case of `energy less than $E$', the additional states don't matter because there aren't so many of them.)   Another criticism of this notion of coarse-grained entropy is that, if the Hamiltonian of which these are energy eigenstates is the Hamiltonian of a theory of everything, then, taken literally, the definition would entail that any system with some given energy $E$ -- be it a black hole, a neutron star or, even, say, a cat  -- would be assigned a coarse grained entropy equal to that of a black hole with the same energy!}) and the resulting notion of entropy partly vague and arbitrary and yet there doesn't seem to be anything subjective or vague or arbitrary about (one quarter of) an area!\footnote{\label{ftntPenQuote} A frequent response to this in the immediate aftermath of Hawking's 1974 announcement is well-exemplified by this quote from the 1979 article \cite{Penrose79} by Penrose

\smallskip

``\textit{Now recall that in the Bekenstein-Hawking formula, an entropy measure is directly put equal to a precise feature of spacetime geometry, namely the surface area of a black hole. Is it that this geometry is now subjective with the implication that all spacetime geometry (and therefore all physics) must in some measure be subjective? Or has the entropy, for a black hole, become objective? If the latter, then may not entropy also become objective in less extreme gravitational situations $\dots$}''

\smallskip

Considerations similar to those in this quote were indeed amongst the inspirations for the present work.}

Thus the second law puzzle acquires a new potency when the closed system in question involves a black hole; in particular, for what we shall call here the (gedanken) \emph{black hole collapse and evaporation system} (see Figure 1) which consists initially of a compact star in an otherwise flat empty universe which collapses to a black hole which subsequently Hawking radiates and eventually disappears, leaving only Hawking radiation streaming away from where the black hole had been. 

Closely related to, but distinct from\footnote{In some of my own past work and also in previous versions of this preprint, I took the phrase `information loss puzzle' to be synonymous with what I call here the `second law puzzle in the case of black holes'.   Here I have altered my terminology so as to make better contact with current usage by other authors.}, the second law puzzle for black holes, is the \emph{black hole information loss puzzle} \cite{HawkingCMP75, HawkingInfoLoss,UnruhWald,Wallace,Maudlin,OkonSudarsky}: How to reconcile the fact that, in a semiclassical understanding, part of the state finishes up on the future singularity and the radiation is predicted to be in a mixed state with the belief/hope that, in a full quantum gravitational understanding, the overall evaporation process will be unitary?

\begin{figure}[h]
\centering
\includegraphics[trim = 6cm 19cm 6cm 4cm, clip]{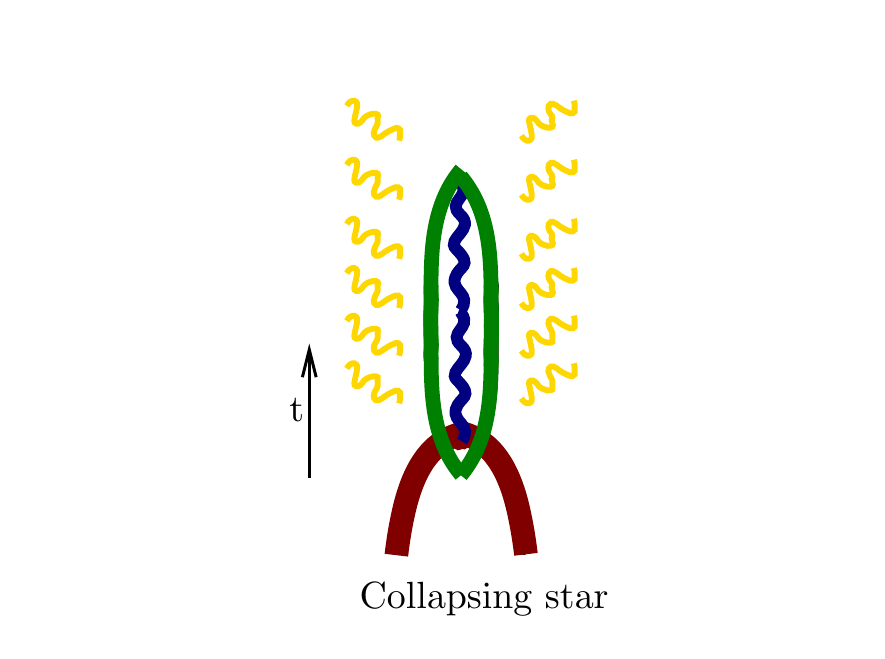}
\caption{A schematic picture of our (spherical) \emph{black hole collapse and evaporation system}.   The red curves are the boundary of the collapsing star, the green curves the horizon, the black wiggly line the singularity and the yellow squiggles the emitted radiation.}
\end{figure}

When this puzzle was first stated by Hawking \cite{HawkingInfoLoss}, he argued that the thermal nature of Hawking radiation was incompatible with unitarity and instead proposed a generalized form of dynamics (or rather of scattering) in terms of a `superscattering operator' acting on density operators.  In later work \cite{Hawking2005, Hawking2015}, partly influenced by progress in string theory, he changed his mind and indeed made some further progress towards showing that unitarity does, after all, hold.   

An influential paper which has led to yet further progress towards resolving the information loss puzzle in favour of unitarity is the 2013 paper of Page \cite{Page2013}.  Page considers a semiclassical description of a black hole collapse and evaporation system for an initially large black hole\footnote{\label{ftntBig} He assumes the only massless particles in nature are photons and gravitons and that one starts with a (star that collapses to a) black hole which is sufficiently large -- and hence sufficiently cold -- that most of the Hawking radiation that gets produced will consist mainly of photons and gravitons.} and, after the formation of the black hole, considers this to consist of two interacting quantum systems, `the black hole' and `the radiation', each spatially localized with a fuzzy boundary in between them at around a few Schwarzschild radii.\footnote{The purpose of the fuzzy boundary is to avoid dealing with the entanglement entropy at the spatial interface between the black hole and its exterior so as, i.a.\ to be able to regard the total entropy as the sum of the black hole entropy and the radiation entropy.   This is to be contrasted with our own proposal  -- discussed below -- according to which the entire physical entropy of the total system is identified with the entanglement entropy between its matter and gravity degrees of freedom.}   (So what Page regards as `the black hole' includes [the part of the radiation field in] a small region outside the horizon.)   He assumes that a freshly formed black hole may be considered to be in a pure state and studies the evolution of the total state consisting initially of such a pure black hole state together with a pure initial state of the radiation field.  (This is not obviously the same as considering an initial total pure state before the stellar collapse to a black hole and it would seem interesting to explore whether this leads to similar results.)  He then argues that for such a state to evolve into a pure state at late times (after the full evaporation of the black hole) so as to be consistent with unitarity, the late emitted quanta (after the `Page time') need to purify (see below for an explanation of this notion) the early emitted quanta.  Recently there has been much work which claims to show, for the black hole collapse and evaporation system, that this indeed happens, as reviewed, for example, by Almheiri et al.\ \cite{AlmheiriEtAl}.  In that work, a number of new types of argument are marshalled which provide evidence that the von Neumann entropy of the radiation indeed goes to zero (as required for unitarity if the black hole evaporates away completely) at late times. 

Let us also mention here that there is a well-known difficulty with the approach of Page and of the work reviewed in \cite{AlmheiriEtAl} which is that, by \emph{monogamy} \cite{monogamy}, the late emitted particles cannot simultaneously purify the early emitted particles and also be entangled (as one might expect them to be on a semiclassical understanding) with their partner-quanta which fell through the horizon.   We shall call this the \emph{firewall puzzle} in view of a much discussed paper \cite{firewall} which emphasizes this difficulty and points out that one way out of it would be to posit the existence of a `firewall' due to the absence of the latter entanglement.  (Statements of essentially the same puzzle also appeared in the earlier papers \cite{BraunsteinEtAl,Mathur}.)   As stated in  \cite{AlmheiriEtAl} it is unclear whether or to what extent the work reviewed in \cite{AlmheiriEtAl} may be considered to fully resolve this firewall puzzle.

The question on which I wish to focus here is what I shall call the \emph{entropy question:}  Taking as given that the underlying dynamics is unitary, can we find a suitable notion of entropy which increases throughout the evaporation process so as to be consistent with the second law? 

Page (and some subsequent authors) address this question by defining the entropy to be the sum of a coarse-grained entropy for the black hole (one definition for which, cf.\ Footnote \ref{ftntCount}, would be the logarithm of the number of black hole states with energy less than the black hole mass) and a similarly-defined coarse-grained entropy for the radiation and arguments have been given that this or related quantities will increase.   
In particular, Wall \cite{Wall2012} has argued for  a  `generalized second law' in the sense that $S_{\rm{black hole}} + S_{\rm{radiation}}$ grows monotonically in time when  $S_{\rm{black hole}}$ is identified with $1/4$ of the horizon area and $S_{\rm{radiation}}$ is defined to be a certain coarse-grained entropy for the Hawking radiation.

Starting in 1998 \cite{Kay1998,KayNewt} I have explored another possible approach to the entropy question.  I took as my starting assumption (see Footnote \ref{ftntPenQuote}) that we need a definition of entropy which is objective and universal.    

Of particular relevance to my train of thought was yet another puzzle about quantum black holes, the \emph{thermal atmosphere puzzle}.   This concerns not the black hole collapse and evaporation system but rather the equilibrium state that we assume \cite{HawkingBHThermo} to be possible if one confines a (spherical, uncharged) black hole, together with its Hawking radiation, to a spherical box. See Figure 2.   We shall call this the (closed) \emph{black hole equilibrium system}.\footnote{\label{ftntBHPic} We refer to \cite{KayEnclosed, KayLupo} where we argue that the best we can do in terms of a classical spacetime picture of the black hole equilibrium system is to picture it (see Figure 2 in \cite{KayEnclosed}) as just the exterior Schwarzschild spacetime but with a non-classically describable edge region, a few Planck lengths thick, around where, classically, the horizon would be.  (This is reminiscent of the firewall of \cite{firewall, BraunsteinEtAl,Mathur} discussed above except the claim here is that it arises as a special feature of equilibrium states enclosed in boxes.)   So Figure 2 here is misleading and we should really think of the gravitational field of the black hole as residing in that edge region outside of/around the horizon.   Let us also mention here, as an aside, that a similar picture is also expected to apply to an AdS black hole and it suggests (see \cite{KayEnclosed, KayOrtiz} and see also \cite{KayEntHol}) that AdS/CFT is not a bijection between the boundary CFT and the full quantum gravity theory in the bulk, but just between the boundary CFT and the matter sector of the bulk, with the gravity sector purifying the matter sector.  It remains, however, an unsolved problem to identify what the matter-gravity split corresponds to in the relevant string theory.}  The puzzle is that there are arguments \cite{PageNewJP,WaldLivRev} that the entropy of such a black hole equilibrium system may be identified with the entropy of the gravitational field of the black hole and there are also arguments \cite{PageNewJP, WaldLivRev} that it may be identified with the entropy of the thermal atmosphere which (in the `inner region' which accounts for most of the entropy -- see below)  consists mostly of matter, albeit a small part of it will consist of gravitons.   And if we were to add those two entropies we would get the wrong answer.  (Twice the correct answer.) 

\begin{figure}[h]
\centering
\includegraphics[scale = 0.7, trim = 6cm 19cm 6cm 5cm, clip]{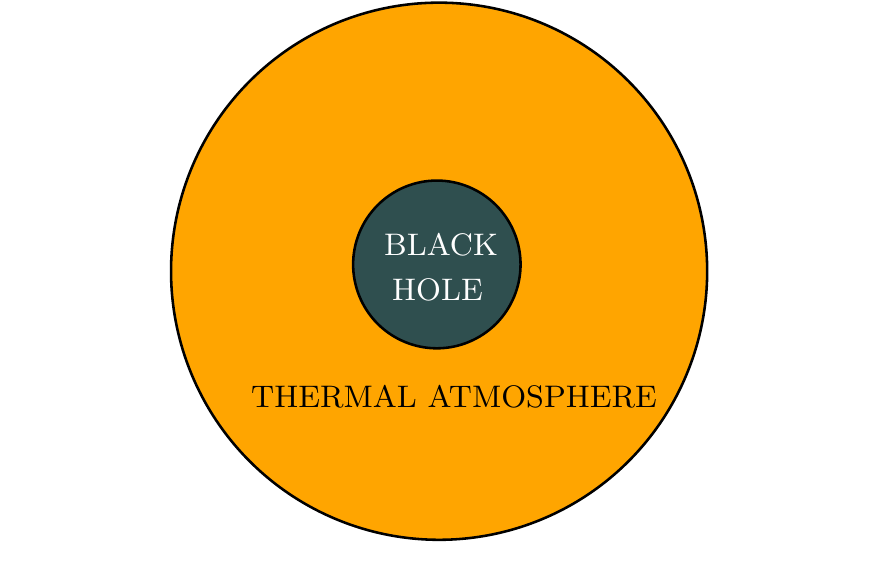}
\caption{A schematic picture of our \emph{black hole equilibrium system}. (See also Footnote \ref{ftntBHPic})}   
\end{figure}

It struck me that this situation is reminiscent of the fact (now very familiar from quantum information theory) that,  if one has a (pure) vector state, $\Psi$, on a bipartite system -- described by a total Hilbert space, $\cal H$, that arises as the tensor product, ${\cal H}_A\otimes{\cal H}_B$ of an `A' Hilbert space and a `B' Hilbert space,  then the partial trace, $\rho_A={\rm Tr}_{{\cal H}_B}(|\Psi\rangle\langle\Psi|)$, of $|\Psi\rangle\langle\Psi|$ over ${\cal H}_B$ (also known as the \emph{reduced state} of the `A' system) is, in general, an impure density operator.   (Similarly with $A \leftrightarrow B$.)  When it is impure, one says that the total state, $\Psi$, is \emph{entangled} and one defines its \emph{A-B entanglement entropy} to be the von Neumann entropy, $S^{\rm{vN}}(\rho_A)=-{\rm tr}(\rho_A\log\rho_A)$ of $\rho_A$ -- which, it turns out, is necessarily equal to the von-Neumann entropy, $S^{\rm{vN}}(\rho_B))$ of $\rho_B$.     One also says that System A \emph{purifies} System B (and vice-versa).\footnote{\label{ftntBipart} Note that, unlike in classical statistical physics, it is perfectly possible therefore to have a pure state on a total system consisting of two subsystems (A and B) each of which, considered separately, are in mixed states.  And in consequence the von Neumann entropy is, in general, very far from being additive.   Also, in view of this fact, this way of obtaining a mixed state from a pure state cannot be regarded as a sort of `coarse-graining'.  (It is unfortunate that in my first account \cite{Kay1998} of my matter gravity entanglement hypothesis, I attempted to explain it in the language of microstates and macrostates and coarse-graining and this aspect of that paper was, in hindsight, unclear and misleading.  This was corrected in my subsequent papers.)}  

The obvious connection to try to make is to identify the von Neumann entropy of $\rho_A$ in the above story with `the entropy of the gravitational field of the black hole' and `the von Neumann entropy of $\rho_B$' with `the entropy of the thermal atmosphere' and the A-B entanglement entropy with the entropy of the entire black hole equilibrium system.   Thus it would seem that the thermal atmosphere puzzle would be resolved if we assume the black hole equilibrium system to be correctly described by a pure total state, $\Psi$, of quantum gravity entangled between the gravitational field of the black hole and its atmosphere in such a way that their reduced states are both approximately Gibbs states at the Hawking temperature. And if we identify the physical entropy of the entire closed system with the total state's entanglement entropy between the gravitational field of the black hole and its thermal atmosphere.   (We remark that, previously, the black hole equilibrium system has been modelled by a total Gibbs state at the Hawking temperature\footnote{It should be mentioned that, aside from our reasons to reject a description in terms of a total Gibbs state in favour of a description in terms of a total pure state, it is anyway problematic to regard a quantum black hole, in the absence of a cosmological constant, as being in a total Gibbs state at the Hawking temperature because it would have a negative specific heat (although this difficulty is absent for certain Schwarzschild AdS black holes \cite{HawkingPage}).  Note anyway that one can infer from the work in \cite{KayThermality,KayMod,KayMore} that the sense in which our reduced states of matter and gravity are expected to be  ``approximately thermal'' also avoids that difficulty.} or [as in Hawking's discussion \cite{HawkingBHThermo} of black hole thermodynamics] in terms of a total microcanonical ensemble.)   As to evidence that the matter-gravity entanglement entropy takes the Hawking value, one can make a semiqualitative argument based on string theory ideas that it goes like the square of the black hole mass (and hence like its area) albeit, along with other string theory approaches to the entropy of Schwarzschild (i.e.\ spherical, uncharged) black holes, one cannot obtain the factor of 1/4.\footnote{\label{ftntString} This argument (see \cite{KayThermality,KayMod,KayMore} or \cite{KayMatGrav} for a review) is a modification of the Horowitz-Polchinski semi-qualitative entropy calculation \cite{HoroPolch} of the entropy of a Schwarzschild black hole, which, in turn, is based on Susskind's picture \cite{Susskind} of the weak string coupling limit (where the string length scale is adjusted to keep Newton's constant of gravitation constant) of a such a black hole as being a long string in Minkowski space.  That argument is subject to the criticisms we gave of  `counting states' in Footnote \ref{ftntCount}.   Instead of looking at the weak string coupling limit of a bare black hole, we look at the weak string coupling limit of the entire black hole equilibrium system (which we take to be a random pure state which arises as a superposition of energy eigenstates with energy in a narrow band around the black hole mass) involving a long string surrounded by a stringy atmosphere of small strings in a box.   We then argue that the matter-gravity entanglement entropy is approximately given by the entanglement entropy of the long string with its stringy atmosphere and, tweaking the procedure in \cite{HoroPolch}, equate this with the entanglement entropy of our black hole equilibrium system when the string length is comparable to the black hole mass.  We find \cite{KayMod,KayMore} that this will go like the square of the black hole mass (while the argument is not subject to the criticisms of Footnote \ref{ftntCount} of `counting of states').  We also hope that e.g.\ the original Strominger-Vafa computation \cite{StromingerVafa} of the black hole entropy of (near) extremal black holes (which is also subject to the `counting states' criticism of Footnote \ref{ftntCount}) might be tweakable in a similar way to be replaced by a calculation of a matter-gravity entanglement entropy.  However this requires one to identify (cf.\ the last remark in Footnote \ref{ftntBHPic}) which degrees of freedom of that string theory correspond to matter and which to gravity.  That is currently still an open problem.}

We want to go further though and to take this as a clue for a universal definition of physical entropy -- i.e.\ a definition for an arbitrary closed system, be it our black hole equilibrium system, be it the black hole collapse and evaporation system, be it the universe!  And we want the notion of physical entropy to make sense, in the case of the black hole collapse and evaporation system, not only during the period of time when there's a black hole present (and emitting radiation) but also before the star collapses and after the black hole has completely evaporated -- and in the case of the universe, at early times soon after the big bang as well as later.  Bearing in mind that, in the case of the black hole equilibrium system, we can roughly equate  `the thermal atmosphere' with `the matter atmosphere' ([near the horizon] gravitons are expected to only be a small part the thermal atmosphere -- we will return to this point later) this suggests that we take our definition for the physical entropy of a closed system to be its \emph{matter-gravity entanglement entropy}.\footnote{\label{ftntOpen} We also have an extension of this hypothesis to include open systems, which also, in a sense, offers a completion of the well-known `environmental decoherence' paradigm.   For this, we refer to \cite[Endnote (xii)]{KayAbyaneh} or \cite{KayEntropy} or \cite{KayRemarks}.}  This in turn suggests that we will need as our fundamental framework a low-energy version of quantum gravity (where low energy means below something like the Planck energy, 10$^{19}$ GeV) for presumably `matter' and `gravity' are not fundamental notions in exact quantum gravity but emerge at such `low' energies.  In consequence, we expect that, on the approach to understanding thermodynamics entailed by our proposed definition for physical entropy, thermodynamics is itself an emergent theory that only makes sense at energies below the Planck energy\footnote{\label{ftntPureGrav} Likewise, our notions of `entropy' and of `thermodynamics' necessarily will not apply to a fictitious universe made out of pure gravity (even though such a universe could contain black holes).  This should not worry us though since, as far as I understand, it is believed that a quantum theory of the pure gravitational field cannot be constructed in a mathematically meaningful way and anyway, we only are interested in having a theory which applies to our universe where, as it happens, there is both gravity and matter.} -- and thus only starts to make sense a few Planck times after the big bang when presumably a sort of phase transition occurs and the distinction between matter and gravity starts to emerge. 

On the assumption that low energy quantum gravity is a quantum theory of a conservative type with a total Hilbert space that arises as the tensor product of a gravity Hilbert space and a matter Hilbert space\footnote{\label{ftntProdPic} It might be objected that the usual constraints of canonical quantum gravity prevent the quantum gravity Hilbert space from arising as a tensor product in this way and prevent the Hamiltonian taking the form (\ref{Ham}) just as, e.g., in quantum electrodynamics in Coulomb gauge, because of the Gauss law constraint, the longitudinal modes of the electromagnetic field are made out of charged matter operators.  However, I have recently shown (see \cite{KayQES,KayQED}) that a (unitarily) equivalent description of QED is possible in which Gauss's law holds as an operator equation which picks out a \emph{physical subspace} of a total Hilbert space which does arise as a tensor product of a charged matter Hilbert space and an electromagnetic field Hilbert space -- albeit all elements of that physical subspace (including the vacuum) have a certain degree of entanglement between charged matter and longitudinal photons.  Work is in progress on generalising this construction and thereby obtaining a Hamiltonian of form (\ref{Ham}) to (first, linearized) quantum gravity.}  and with a time-evolution, $U(t) = e^{-iHt}$ which is unitary\footnote{\label{ftntTime} From what we know about the nature of `time' in general, or even special, relativity, to talk about `time-evolution' in quantum gravity in such a na\"ive way raises difficult questions to which, we must admit, in the case of our black hole collapse and evaporation system, we are presently unable to provide a clear answer except in a hand-waving sense or in the limit of weak string coupling and large string length -- see Footnote \ref{ftntString} and \cite{KayThermality,KayMod,KayMore,KayMatGrav}.  Let us note though that in the case of our black hole equilibrium system, there is an obvious notion of time with respect to which the system is static.  Also, in the case of the universe, it is natural to identify `time' here with cosmological time.} we then, arguably have a resolution to the second law puzzle since we now have a notion of physical entropy -- namely matter-gravity entanglement entropy -- which is not subjective and \emph{not} a unitary invariant.  If we make the additional assumption that the total Hamiltonian, $H_{\rm{total}}$, arises as the sum 
\begin{equation}
\label{Ham}
H_{\rm{total}} = H_{\rm{gravity}} + H_{\rm{matter}} + H_{\rm{interaction}},
\end{equation}
of a gravity Hamiltonian, a matter Hamiltonian and an interaction term, and if we further assume\footnote{It is routine in many sorts of explanations for entropy increase, to assume that the initial state of the universe (and hence also of branch subsystems -- see \cite{Reich,Davies} -- of the universe) has a low entropy and this assumption, for us, of course means a low degree of matter-gravity entanglement.   At present we just add this to our other assumptions but there is a hope that it could emerge from some future more fundamental theory of the initial state.} that the initial state (in the case of the universe, a few Planck times after the big bang) is unentangled between matter and gravity (or rather -- cf.\ Footnote \ref{ftntProdPic} -- has a low degree of matter-gravity entanglement) then it seems reasonable to expect that the degree of entanglement, and therefore the physical entropy as we define it, will increase because of the interaction.   (If our systems had a finite number of degrees of freedom, one might anticipate Poincar\'e recurrence times when the entropy returned to a low value, but if the number of degrees of freedom is large [or infinite] that might well not happen for a very long time [or ever].)  Thus our hypotheses seem capable of offering a resolution to the second law puzzle, and of offering a plausible mechanism for why the second law holds -- i.e.\ why the physical entropy of a closed system increases with time.  We shall call this our \emph{general argument for entropy increase}.   At the very least, we can say that with our proposed definition for entropy, the question of whether the second law holds becomes an objective question whose answer depends on the initial state and the (low energy) quantum gravitational Hamiltonian.

We will not pursue that question further in its full generality here.   Rather we shall attempt to  understand, in the case of the black hole collapse and evaporation system, whether (and if so, how) the entropy as we define it, will increase in terms of the details of the collapse and evaporation process (while unitarity is maintained).   Here we remark that, on our matter-gravity entanglement hypothesis, it seems that any configuration of ordinary matter will have a nonzero entropy, albeit this is expected to be small unless the system is highly relativistic and/or involves strong gravitational fields.  (See \cite{KayNewt,KayAbyaneh} where we also give some evidence that the entropy will be lower e.g.\ for a gas and higher when matter is in a more condensed state.)  Thus our collapsing star will already have a small nonzero entropy and presumably this will increase rapidly to the Hawking value as the black hole forms, presumably because of a large degree of matter-gravity entanglement happening around where, in a classical description, a horizon forms.    But also, if the second law is to hold, then it must be that, after the full evaporation of the black hole, the state of Hawking radiation that remains, streaming outwards from the centre where the black hole had been, has a physical entropy even greater than the entropy that the black hole had when it was freshly formed (i.e.\ as we know from Hawking, one quarter of the area of its horizon).  How can that be, if we define physical entropy to be matter-gravity entanglement entropy?   If the initial black hole was large enough and assuming that the only massless particles in nature are photons and gravitons (see Footnote \ref{ftntBig}) then (cf.\ \cite{Page2013}) the late time Hawking radiation will mainly consist just of photons and gravitons.   For this to have the necessary big entropy as we define it (and for the total state to remain pure!) it presumably must be that the photons and gravitons are highly entangled with one another (while in a total pure state).  (Here we are assuming that matter-gravity entanglement entropy amounts, in this case, to photon-graviton entanglement entropy.)

Now there immediately appears a difficulty.   According to the 1976 Hawking-effect calculations of Page \cite{Page1976, Page1976II}, because of their higher spin, fewer gravitons will be emitted than photons.  Moreover, even if one purifies as many photons as one can with the fewer gravitons predicted to be emitted, one may easily infer from the quantitative results of \cite{Page1976II} (as cited in \cite{Page2013} -- see also \cite{Page1983} and \cite{PageNewJP})  that the resulting matter-gravity entanglement entropy would fall well short of the entropy of the freshly formed black hole.  (In fact, using the results in \cite{Page1976II,Page2013} one finds that one would only obtain $0.15198$ times the entropy of the freshly formed black hole.)  And there is also no mechanism for any such purification in those 1976 calculations.

However there seem to be a couple of important possible loopholes.   Those calculations in \cite{Page1976,Page1976II}, as indeed the original black hole evaporation calculation \cite{Hawking75} of Hawking, are done in the framework of quantum field theory in curved spacetime and it may well be that, at least after a black hole has been radiating for some time, the quantum fluctuations around the horizon build up and render that approximation to quantum gravity (and also semiclassical gravity where one incorporates a backreaction of the expectation value of the stress energy tensor) inapplicable.\footnote{\label{ftntLate} We remark that some authors assume that semiclassical gravity (in one form or another) will, instead, hold until the final stages of black hole evaporation when the black hole's size becomes comparable to the Planck length.  However we are unaware of any strong arguments that any version of semiclassical gravity can be trusted for so long.} 

Secondly, the work in \cite{Page1976,Page1976II} neglects photon-graviton interactions.   Let us temporarily leave aside our black hole collapse and evaporation system and look in some detail at equilibrium states of quantum gravity confined to a gedanken spherical box of volume $V$.  We model these as random pure total states with energy in a narrow band around $E$\footnote{\label{ftntBand} By this we mean they arise as superpositions of energy eigenstates with energies in a narrow band around $E$.} Although we are dealing with randomly chosen pure total states and Hawking \cite{HawkingBHThermo} studied a microcanonical ensemble, we expect (see \cite{KayThermality,KayMod,KayMore}) that the broad features of the thermodynamical behaviour that is predicted on our assumptions will be qualitatively similar to those described in \cite[Page 195]{HawkingBHThermo}.   (In the terminology of \cite{KayThermality,KayMod} we are replacing the `traditional' by the `modern' explanation of the origin of thermodynamic behaviour.)  If $V$ is sufficiently large, we expect that the box would contain just radiation and no black hole and for that radiation to mainly consist just of massless particles -- i.e.\ photons and gravitons.  But now we would expect the total state to involve equal numbers of photons and gravitons and moreover (on our assumption of a total pure state) for that state to be highly entangled between the photons and gravitons.  After all, all that seems to matter here is that we have a total pure state of two species of massless boson, each with two helicity states and weakly coupled to one another.   And by arguments similar to those in \cite{KayThermality} in such a situation (two weakly coupled systems with the same density of eigenvalues in a random pure total state with energy in a narrow band) one expects the state to be highly entangled.

Next consider a suitably smaller volume $V$ (but not so small as for its radius to be within the Schwarzschild radius for energy $E$) then (cf.\ again \cite[Page 195]{HawkingBHThermo}) we would expect a random energy eigenstate with energy around $E$ (in the sense made precise in Footnote \ref{ftntBand}) to very probably consist of a (spherical) black hole (let us assume, for simplicity, located centrally and ignore its centre of mass motion) surrounded by an atmosphere, thus providing a model for what we called above our `black hole equilibrium system'.   Further, we would expect this atmosphere to roughly consist of an inner region, where $r/2M-1 \ll 1$, where, because of the blueshift, all matter degrees of freedom are effectively massless and (cf.\ 't Hooft's brick wall discussion \cite{tHooft, MukohyamaIsrael} and our discussion of the thermal atmosphere puzzle above as well as Footnote \ref{ftntBHPic}) that it is the entanglement of the matter fields in this region with the gravitational degrees of freedom that will account for most of the entropy, and an outer region consisting just of massless particles -- i.e.\ just photons and gravitons.   And, similarly to in the case where a black hole is absent, we would expect those outer photons and gravitons to be in roughly equal numbers and highly entangled with one another, albeit their contribution  the overall (matter-gravity entanglement) entropy will be small compared to that of the inner region.  (It also seems reasonable to expect that the entropy of the inner and outer regions will roughly add up to the total entropy.)   It also seems plausible that those outer photons will (consistently with monogamy \cite{monogamy}) not be strongly entangled with the gravitational field of the inner region.  Now, suppose we were to make a small hole in the box in the latter situation -- or rather, to preserve spherical symmetry, suppose we were to make the box slightly permeable.   Presumably what would come out would be radiation consisting of photons and gravitons --  in roughly equal numbers and highly entangled with one another, while the escaped photons and gravitons would be continually replenished (at the expense of the inner region) as the system inside the box strives to return to equilibrium.  

So, if we start with a black hole in a spherical box in equilibrium and then permit it to evaporate slowly by making the box slightly permeable, it seems possible that what we will be left with after it has fully evaporated, is a pure total state of radiation consisting of photons and gravitons --  in roughly equal numbers and highly entangled with one another.   It seems possible that this might have a (matter-gravity entanglement) entropy greater than that of the freshly formed black hole although whether it does remains an open question still to be investigated.    Let us say in support of this possibility that  it seems possible that the partial state of the photons obtained by tracing over the gravitons (whose von Neumann entropy is of course the photon-graviton entanglement entropy) might not be too different  from the photon component of Hawking radiation calculated using quantum field theory in curved spacetime for a semiclassical model of an evaporating black hole and let us note that, say for an initial Schwarzschild black hole, it easily follows from Page's 1976 calculations \cite{Page1976II,Page1983,Page2013} that the latter would have a (von Neumann) entropy 1.33274 times the entropy of the freshly formed black hole.  

It also seems plausible that the same will be true if the initial state inside the slightly permeable box is not an equilibrium state but the state of a black hole freshly formed by collapse and then enclosed in the box since we would anyway expect that such a system would strive to surround itself with an atmosphere by Hawking radiating.    

Returning to the standard black hole collapse and evaporation system, let us first reiterate that, if we accept our general argument for entropy increase, we would still expect that, at the same low-energy quantum gravity level of description and again due to photon-graviton interactions (which are neglected in a quantum field theory in curved spacetime or semiclassical treatment) the Hawking radiation would still consist, at late times, of a pure state of entangled photons and gravitons with an entropy greater than that of the freshly formed black hole.  To attempt to give a detailed physical understanding of how this might come about seems more difficult and subject to even more uncertainties than for the above-discussed case when the black hole is enclosed by a slightly permeable box.   But we would expect (see the first diagram in Figure 3) that (highly blueshifted) Hawking-emitted photons emerging from just outside the horizon will radiate (soft) gravitons due to the fact that the photons will be freely falling in the gravitational field of (what remains of) the black hole. (The photons will also redshift as they climb out of the gravitational potential.)    One expects such radiation to happen just as one expects that an electron freely falling in the gravitational field of a black hole or of a star will radiate soft photons.\footnote{It has sometimes been argued by an application of the equivalence principle that a charged particle freely falling in a gravitational field will not radiate but there are strong arguments (see e.g.\ \cite{UmGillies}) that such an application of the equivalence principle is illegitimate.  And, anyway (see e.g.\ \cite{MorBerCrisp} and references therein) we know e.g.\ that a charged particle orbiting a black hole will emit synchrotron radiation and also indeed that an orbiting massive particle will radiate gravitons.}   Moreover, one expects that as it radiates gravitons, each Hawking photon will become entangled with them, just as \cite{InfraQI,DresInfraQI} (and see \cite{PrabSatWald} for the generalization to massless particles) an outgoing electron in M{\o}ller scattering will radiate, and become entangled with, soft photons.   However, it remains to be investigated whether, in the course of its travel away from the horizon, the resulting degree of entanglement of a photon with its radiated gravitons will be high.  One also, of course, needs to consider what happens to Hawking-emitted gravitons.  One expects (see again Figure 3) Feynman-like diagrams where (in the presence of the gravitational background) such a graviton can produce a pair of photons (at least one of which will be comparably `hard' to the graviton and would then in its turn emit more soft gravitons etc.) and higher order in Newton's constant diagrams where the gravitons continue on their way after emitting even numbers of soft photons as in the third diagram of Figure 3.  Again it remains to be investigated whether the resulting degree of graviton-photon entanglement is high.

Let us remark here that, for an initially non-large black hole, or after an initially large black hole has reduced in size so much that Hawking radiation of massive particles becomes important, one expects that the same mechanism, i.e.\ that the massive particles radiate soft gravitons and become entanglement with them, would apply.

\begin{figure}[h]
\centering
\includegraphics[scale=0.20, bb=-2in 2in 14in 10in, clip]{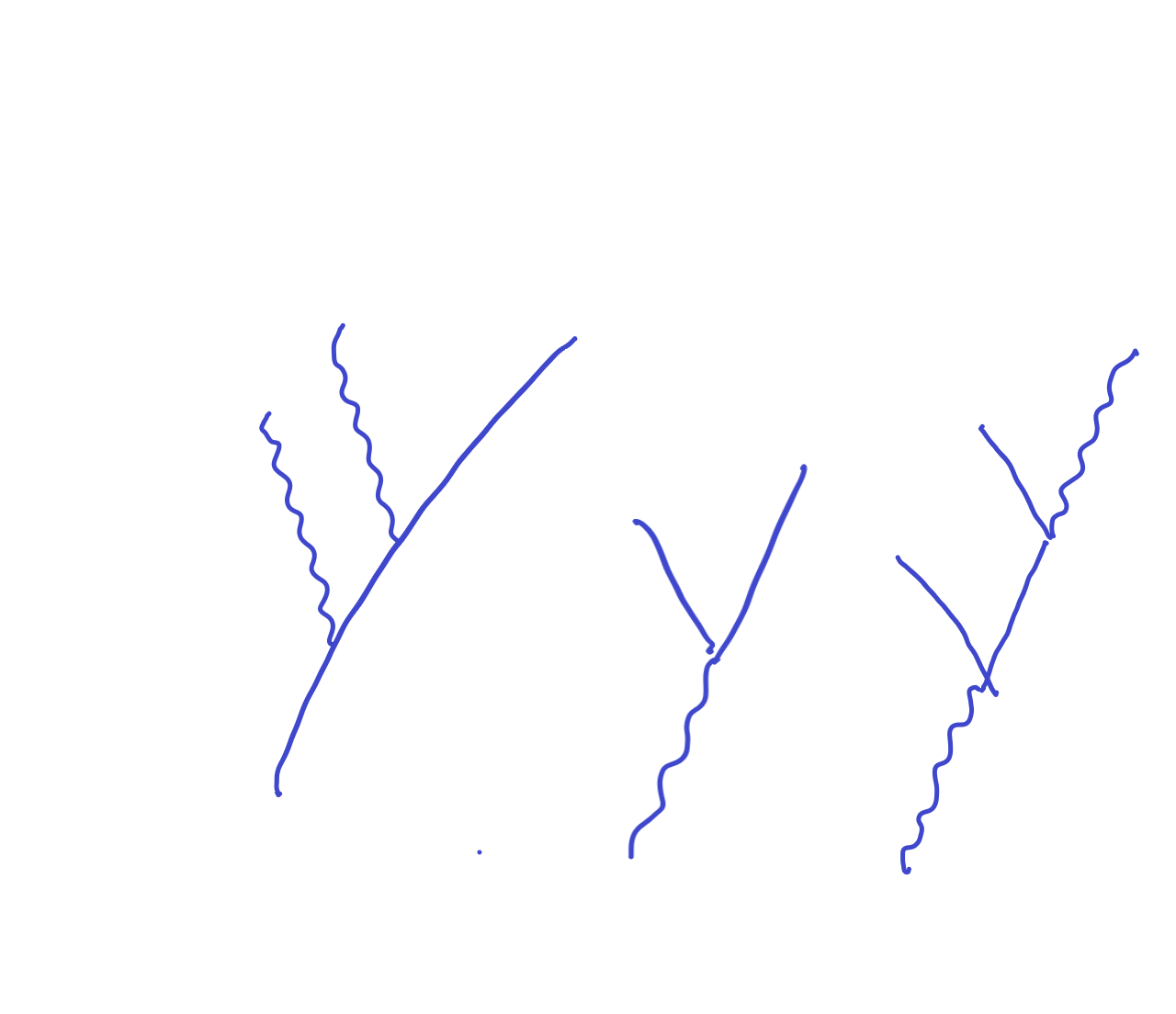}
\caption{Feynman-like diagrams (straight line $=$ photon, wiggly line $=$ graviton) showing a photon emerging from near the horizon emitting (soft) gravitons; a (hard) graviton emitting (say) a hard and a soft photon; and a graviton converting to a photon and then back to a graviton after emitting two (soft) photons}
\end{figure}

There is an important caveat here that we must address.   
Even though, in this scenario, Hawking-emitted photons are themselves emitting, and entangling themselves with, gravitons (and there is a corresponding story for Hawking-emitted gravitons as sketched above) and thereby, insofar as a semiclassical picture of black hole evaporation is valid, will get less entangled with their partner photons etc.\ that had fallen through the horizon, one would expect, on that semiclassical picture, the whole subsystem consisting, say, of a single Hawking-emitted photon together with all the soft gravitons that it itself radiates to remain just as entangled with that partner photon as the freshly emitted Hawking photon had been.\footnote{For this reason and contrary to what we suggested in previous versions of this preprint, taking into account photon-graviton interactions would not seem to improve the prospects for resolving the firewall puzzle.  (I thank Michael Kay and Daniel Sudarsky for discussions on this point.) However it could of course be (see Footnote \ref{ftntLate}) that at non-early times Hawking evaporation cannot be understood in semiclassical terms in which case this may not necessarily trouble us.}   (For the degree of entanglement to change would require an interaction between that whole subsystem and the partner photon.)  Thus one would still presumably need to invoke a result along the lines of the results reported in \cite{AlmheiriEtAl} to understand how (as previously predicted by Page) the late emitted quanta (now presumably together with their own emitted soft gravitons etc.) would purify such subsystems.

In conclusion, we have sketched possible mechanisms based on photon-graviton interactions (and on interactions of other matter particles with gravitons) for how entropy as we define it  (i.e.\ matter-gravity entanglement entropy) might increase throughout the formation and subsequent evaporation of a black hole, thereby also offering support for our above general argument for entropy increase in the case of our black hole collapse and evaporation system.  If the freshly formed black hole is enclosed in a slightly permeable box,  we argued that this is possible by relating the late-time state of Hawking radiation to the state of the outer part of the thermal atmosphere in the black hole equilibrium system.   In the absence of such a box, we pointed out that one would expect i.a.\ Hawking-emitted photons themselves to radiate gravitons and become entangled with them.  More work is needed to test if these possibilities actually hold, in particular whether the latter degree of entanglement will be high enough for the entropy to always increase.   If it will be, then one would expect the late-time state of Hawking radiation to largely consist of photons highly entangled with, and purified by, gravitons.

\end{document}